\documentclass[twocolumn]{aastex631}
\received{x}
\revised{x}
\accepted{x}
\submitjournal{ApJL}
\usepackage{fancyvrb} 
\usepackage{breakurl}
\usepackage{multirow}
\usepackage{xcolor}
\VerbatimFootnotes    

\begin{document}

\newcommand{\vdag}{(v)^\dagger}
\newcommand\aastex{AAS\TeX}
\newcommand\latex{La\TeX}

\shorttitle{Updated Multimessenger Picture of J1048+7143}
\shortauthors{Kun et al.}


\title{Follow-up on the Supermassive Black Hole Binary Candidate J1048+7143: Successful Prediction of the Next Gamma-ray Flare and Refined Binary Parameters in the Framework of Jet Precession Model}

\correspondingauthor{Emma Kun}
\email{emma@tp4.ruhr-uni-bochum.de}

\author[0000-0003-2769-3591]{Emma Kun}
\affiliation{Astronomical Institute, Faculty for Physics \& Astronomy, Ruhr University Bochum, 44780 Bochum, Germany}
\affiliation{Theoretical Physics IV: Plasma-Astroparticle Physics, Faculty for Physics \& Astronomy, Ruhr University Bochum, 44780 Bochum, Germany}
\affiliation{Ruhr Astroparticle And Plasma Physics Center (RAPP Center), Ruhr-Universit\"at Bochum, 44780 Bochum, Germany}
\affiliation{Konkoly Observatory, HUN-REN Research Centre for Astronomy and Earth Sciences, Konkoly Thege Miklós \'ut 15-17, H-1121 Budapest, Hungary}
\affiliation{CSFK, MTA Centre of Excellence, Konkoly Thege Miklós \'ut 15-17, H-1121 Budapest, Hungary}

\author[0000-0001-5180-2845]{Ilja Jaroschewski}
\affiliation{Theoretical Physics IV: Plasma-Astroparticle Physics, Faculty for Physics \& Astronomy, Ruhr University Bochum, 44780 Bochum, Germany}
\affiliation{Ruhr Astroparticle And Plasma Physics Center (RAPP Center), Ruhr-Universit\"at Bochum, 44780 Bochum, Germany}

\author[0000-0002-1748-7367]{Julia Becker Tjus}
\affiliation{Theoretical Physics IV: Plasma-Astroparticle Physics, Faculty for Physics \& Astronomy, Ruhr University Bochum, 44780 Bochum, Germany}
\affiliation{Ruhr Astroparticle And Plasma Physics Center (RAPP Center), Ruhr-Universit\"at Bochum, 44780 Bochum, Germany}
\affiliation{Department of Space, Earth and Environment, Chalmers University of Technology, 412 96 Gothenburg, Sweden}

\author[0000-0001-9240-6734]{Silke Britzen}
\affiliation{Max-Panck-Institut für Radioastronomie, Auf dem Hügel 69, 53121 Bonn, Germany}

\author[0000-0003-3079-1889]{S\'andor Frey}
\affiliation{Konkoly Observatory, HUN-REN Research Centre for Astronomy and Earth Sciences, Konkoly Thege Miklós \'ut 15-17, H-1121 Budapest, Hungary}
\affiliation{CSFK, MTA Centre of Excellence, Konkoly Thege Miklós \'ut 15-17, H-1121 Budapest, Hungary}
\affiliation{Institute of Physics and Astronomy, ELTE E\"{o}tv\"{o}s Lor\'{a}nd University, P\'{a}zm\'{a}ny P\'{e}ter s\'{e}t\'{a}ny 1/A, H-1117 Budapest, Hungary}

\author[0000-0003-1020-1597]{Krisztina \'Eva Gab\'anyi}
\affiliation{Department of Astronomy, Institute of Physics and Astronomy, ELTE E\"{o}tv\"{o}s Lor\'{a}nd University, P\'{a}zm\'{a}ny P\'{e}ter s\'{e}t\'{a}ny 1/A, H-1117 Budapest, Hungary}
\affiliation{HUN-REN--ELTE Extragalactic Astrophysics Research Group, ELTE E\"{o}tv\"{o}s Lor\'{a}nd University, P\'{a}zm\'{a}ny P\'{e}ter s\'{e}t\'{a}ny 1/A, H-1117 Budapest, Hungary}
\affiliation{Konkoly Observatory, HUN-REN Research Centre for Astronomy and Earth Sciences, Konkoly Thege Miklós \'ut 15-17, H-1121 Budapest, Hungary}
\affiliation{CSFK, MTA Centre of Excellence, Konkoly Thege Miklós \'ut 15-17, H-1121 Budapest, Hungary}

\author[0000-0003-0721-5509]{Lang Cui}
\affiliation{Xinjiang Astronomical Observatory, Chinese Academy of Sciences, 150 Science 1-Street, Urumqi 830011, China}

\author[0000-0002-9373-3865]{Xin Wang}
\affiliation{Xinjiang Astronomical Observatory, Chinese Academy of Sciences, 150 Science 1-Street, Urumqi 830011, China}

\author{Yuling Shen}
\affiliation{Xinjiang Astronomical Observatory, Chinese Academy of Sciences, 150 Science 1-Street, Urumqi 830011, China}

\begin{abstract}
Analyzing single-dish and VLBI radio, as well as \textit{Fermi}-LAT $\gamma$-ray observations, we explained the three major $\gamma$-ray flares in the $\gamma$-ray light curve of FSRQ J1048+7143 with the spin--orbit precession of the dominant mass black hole in a supermassive black hole binary system. Here, we report on the detection of a fourth $\gamma$-ray flare from J1048+7143, appearing in the time interval which was predicted in our previous work. Including this new flare, we constrained the mass ratio into a narrow range of $0.062<q<0.088$, and consequently we were able to further constrain the parameters of the hypothetical supermassive binary black hole at the heart of J1048+7143. We predict the occurrence of the fifth major $\gamma$-ray flare that would appear only if the jet will still lay close to our line sight. The fourth major $\gamma$-ray flare also shows the two-subflare structure, further strengthening our scenario in which the occurrence of the subflares is the signature of the precession of a spine--sheath jet structure that quasi-periodically interacts with a proton target, e.g. clouds in the broad-line region.
\end{abstract}


\section{Introduction}
\label{section:intro}

In the recent years, multimessenger astronomy became one of the most rapidly evolving fields in astronomy. As being the synergy of coordinated observations targeting all of the four extragalactic messengers, such as the electromagnetic (EM) radiation, cosmic rays, neutrinos, and gravitational waves (GWs), multimessenger astronomy is a very useful tool for studying the most energetic phenomena of the Universe, such as the merging of supermassive black holes (SMBHs), see e.g.\ \cite{2008PhR...458..173B}.

In 2011, the IceCube Collaboration detected the high-energy neutrino flux with cosmic origin for the first time ever \citep{2013Sci...342E...1I}. In 2016, the LIGO Scientific Collaboration and Virgo Collaboration announced the twofold direct detection of gravitational waves (GWs) in both Advanced LIGO observatories in 2015 \citep{2016PhRvL.116f1102A}. In 2017, the GW signal of two colliding neutron stars, as well as a short $\gamma$-ray burst signal have been also detected \citep{2017PhRvL.119p1101A,2017ApJ...848L..13A}. Patterns of high-energy particles sources are thoroughly discussed in the literature \citep{IC2020tenyears,Franckowiak2020,Kun2022MW,Novikova2023}, however, there are only two direct neutrino-source associations yet \citep{ICTXS2018a,ngc1068_2022}. The simultaneous observation of GWs and high-energy neutrinos still lies ahead. 

Since neutrinos interact only weakly with matter, they are able to pinpoint cosmic high-energy particle accelerators in the sky that would be otherwise hidden because ultra-high energy cosmic rays (UHECRs) scatter due to the Galactic and intergalactic fields \citep{Dermer2009}, while their energy is strongly attenuated beyond the Greisen--Zatsepin--Kuzmin (GZK) horizon \citep[e.g.][]{Stanev2000}. High-energy $\gamma$-photons lose also their energy on their way to Earth in interactions with the photons of the Cosmic Microwave Background (CMB). The cosmic neutrinos are indeed the key messengers to explore the Universe where it is opaque to the cosmic rays and photons. Violent phenomena in the distant Universe, for example merging SMBHs in the young Universe, can be observed through the observation window of GWs and/or neutrinos only. Based on the potential detection of the gravitational wave background by the NANOGrav Collaboration \citep{2023ApJ...951L...8A}, supermassive binary black hole (SMBBH) mergers should happen more often than previously assumed. Interpreting the X-shaped jets of radio galaxies as relics of jet precession, which is a plausible assumption e.g. based on their mass excess \citep[see][]{2012RAA....12..127G}, one can argue that there are more such mergers potentially observed. The frequency at which the SMBH mergers happen is not quite clear though, and potential periodic neutrino sources might help understand the mergers and their rate better.

In \citet{Kun2022a}, we analyzed and modeled the \textit{Fermi} Large Area Telescope (LAT) light curve of the blazar J1048+7143, and concluded that the multiwavelength behavior of this source is compatible with the spin--orbit precession of a SMBBH \citep{GerPLB2009} in the heart of the galaxy. This active galactic nucleus (AGN) indeed belongs to a group of blazars which reveal precession-induced variability \citep[see][and references therein]{britzen23}. Generalizing the model of \citet{deBruijn2020}, we predicted the GW signal of this blazar. We also predicted the appearance time of a fourth $\gamma$-ray flare. This model was already successfully applied to the multi-epoch neutrino observations from the direction of TXS 0506+056 \citep{ICTXS2018a,ICTXS2018b} by \citet{Tjus2022}. \citet{britzen19} also proposed a SMBBH for the same source. \citet{deBruijn2020} predicted a neutrino episode for TXS 0506+056, which was later confirmed by the detection of a high-energy neutrino by the Antarctic IceCube Neutrino Detector from the approximate direction of this blazar. We note that this neutrino only skimmed the edge of the detector, therefore its uncertainty (90\% containment) is a relatively large $\sim3.58\degr$ \citep[the neutrino arrived with the energy of $\sim 170$ TeV and the event has 
a signalness of 42\%,][]{IC220918A_gcn1}. In this paper, we extend the $\gamma$-ray light curve of J1048+743 with analyzing the new \textit{Fermi}-LAT data, and indeed find the predicted fourth flare. Using the extended $\gamma$-ray light curve, we fine-tune the binary parameters that are consistent with the observations within our model.

\section{The Updated $\gamma$-ray Light Curve of J1048+7143}
\label{sec:gamma}

\begin{deluxetable*}{ccccccccc}
\tablecaption{Exponential fitting of the $\gamma$-ray light curve of 4FGL~J1048.4+7143. Column (1) sets the fitted quantity ($a$ measures the height of the respective flare, $b$ the time location of its peak, and $c$ its slope). The numbered columns (2)--(7) contain the values of the parameters of the exponential functions fitted to the respective flares, where $F_{i,j}$ means the $j$-th subflare ($j$=1 or $j$=2) of the $i$-th main flare (from $1$ to $4$). The fitted baseline is $a_0=(0.12\pm0.02) \times 10^{-7} \mathrm{ph}~\mathrm{cm}^{-2} \mathrm{s}^{-1}$. 
\label{table:exp_flares}}
\tablewidth{0pt}
\tablehead{
\colhead{Parameter} & \colhead{$F_{1,1}$} & \colhead{$F_{1,2}$} & \colhead{$F_{2,1}$} & \colhead{$F_{2,2}$} & \colhead{$F_{3,1}$} & \colhead{$F_{3,2}$} &\colhead{$F_{4,1}$} & \colhead{$F_{4,2}$}}
\startdata
\multicolumn{1}{p{2.8cm}}{\centering $a$\,($10^{-7}\mathrm{ph}\,\mathrm{cm}^{-2}\,\mathrm{s}^{-1}$)} & $1.85\pm0.37$ & $3.35\pm0.58$ & $2.57\pm0.64$ & $3.95\pm0.55$ & $4.00\pm0.54$ & $3.63\pm0.58$ & $1.05\pm0.53$ & $2.18\pm0.34$ \\
$b$ (MJD, days) & $56378\pm16$ & $56710\pm9$ & $57577\pm15$ & $57801\pm10$ & $58760\pm7$ & $58957\pm7$ & $59555\pm18$ & $60033\pm22$\\
$c$ (days) & $113\pm31$ & $71\pm16$ & $77\pm21$ & $88\pm16$ & $73\pm14$ & $58\pm12$ & $58\pm47$ & $162\pm50$\\
\enddata
\end{deluxetable*}
\begin{figure*}
    \centering
    \includegraphics[angle=0,scale=0.465]{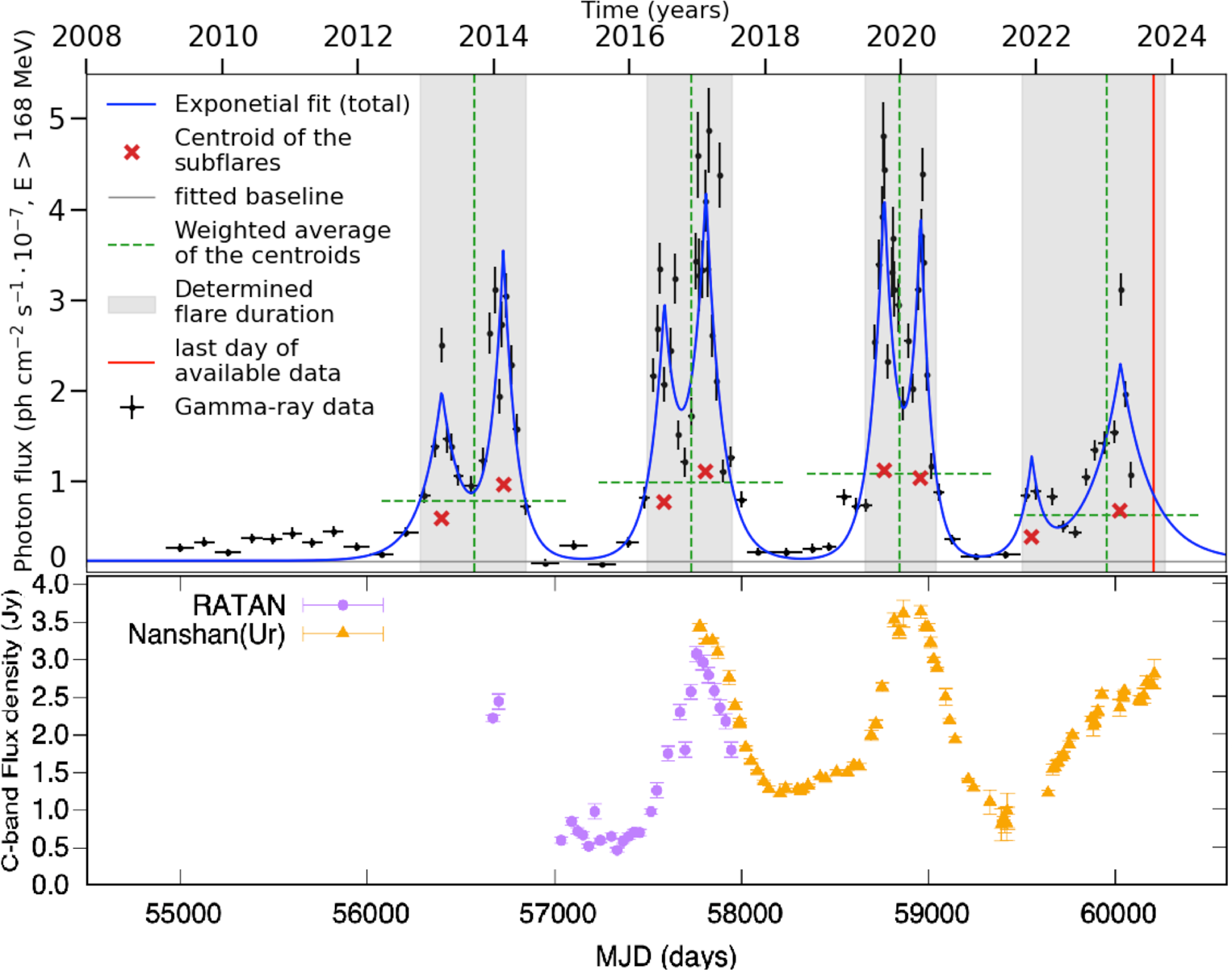}
    \caption{Gamma-ray and radio flares of J1048+7143. Top: The new exponential fit of the $\gamma$-ray light curve of J1048+7143 with 5\% adaptive binning (best-fit parameters are shown in Table~\ref{table:exp_flares}). The sum of the contributions from individual sub-flares is displayed as the blue continuous line. Application of the centroid method is indicated via the red crosses. Their weighted average determine the centers of main flares as green dashed line. Intersections of these lines with the blue lines indicate the duration of the flares, illustrated as gray areas. The red continuous line shows the latest data point available in the \textit{Fermi}-LAT Light Curve Repository (2023 September 22 or MJD 60209). Bottom: The extended radio flux density curve of J1048+7143, obtained with the RATAN600 and  Nanshan (Urumqi) radio telescopes at $4.8$~GHz \citep[see details in][]{Kun2022a}. }
    \label{fig:centroid_Method}
\end{figure*}

We obtained archival data taken with the LAT instrument onboard the \textit{Fermi} Gamma-ray Space Telescope to measure the $\gamma$-ray flux of 4FGL~J1048.4+7143 \citep[associated with J1048+7143,][]{Fermi4fgldr12020}, located at the sky coordinates of right ascension $\mathrm{RA_{J2000}}=162\fdg1067$ and declination $\mathrm{DEC_{J2000}}=71\fdg7297$. The region of interest (ROI) with a $15\degr$ radius was centered at the J2000 sky coordinates of 4FGL~J1048.4+7143. We extended our previous analysis published in \cite{Kun2022a} by adding \textit{Fermi}-LAT data in the time range from 2022 Mar 14 to 2023 Jun 4 (MJD 59635--60099) in the energy range $100\,\mathrm{MeV}-800\,\mathrm{GeV}$. Now our analysis covers almost $15$ years between 2008 Aug 4 and to 2023 Jun 4.
 
 \begin{deluxetable*}{cccccccc}
\tablecaption{Flare characteristics using the centroid method. 
\label{table:flare_characteristics}}
\tablewidth{0pt}
\tablehead{
\colhead{Parameter} & \colhead{F1} & \colhead{$P_{1\to2}$} & \colhead{F2} & \colhead{$P_{2\to3}$} & \colhead{F3} & \colhead{$P_{3\to4}$} & \colhead{F4}}
\startdata
Flare center (MJD) & $56554 \pm 38$  && $57721 \pm 24$  && $58842 \pm 17$ && $59958 \pm 59$\\
Flare duration (d) & $568 \pm 75$  && $450 \pm 60$  && $385 \pm 36$ && $756 \pm 113$  \\
\multicolumn{1}{p{2.8cm}}{\centering Time till next flare center (yr)} & & \multirow{2}{*}{$3.20 \pm 0.12$} && \multirow{2}{*}{$3.07 \pm 0.08$} && \multirow{2}{*}{$3.06 \pm 0.17$} & \\
\enddata
\end{deluxetable*}


We repeated the same binned likelihood analysis of \textit{Fermi}-LAT data for the time range MJD 59635--60099 as we did earlier \citep{Kun2022a}. Technical details of the analysis chain can be found in that work. The binning of the light curve and the optimum energy were constrained using adaptive binning with 5\% threshold of relative flux error. In the resulting light curve, shown in Fig.~\ref{fig:centroid_Method}, we see a new major flare starting in the beginning of 2021 (after MJD 59200), in addition to the previously reported three flares \citep{Kun2022a}. This 4th major flare also shows a double sub-flare structure similar to the three major flares already reported in \citet{Kun2022a}. We refitted the light curve with two-sided exponential functions in the form of
\begin{equation}
    F_{i,j}(t)=a_0 + a_{i,j} \times \exp\left[{\frac{-|t-b_{i,j}|}{c_{i,j}}}\right],
\end{equation}
where $t$ is time, the index $i$ runs from $1$ to $4$ for the four main flares, $j$ is 1 or 2 for the subflares of the $i$th main flare, $a_{i,j}$ the height, $b_{i,j}$ the time location of the peak, and $c_{i,j}$ the slope of the exponential function. Since in the quiescent phase the $\gamma$-ray flux did not vanish, we also fitted a constant baseline (represented by $a_0$). We present the resulting fit parameters in Table~\ref{table:exp_flares}. We derived the center time of each of the four major flares in the extended light curve using the centroid method described in \citet{Kun2022a}. We present the flare characteristics using the centroid method in Table~\ref{table:flare_characteristics} and show the centroids in Fig.~\ref{fig:centroid_Method} as the cross of the green dashed lines. The intersection of the horizontal green dashed line with the exponential fit (blue continuous line) of the respective main flare indicates the start and end of the flare. The duration is thus seen as the gray shaded area and given with error bars in Table~\ref{table:exp_flares}. Its uncertainty is determined using the Markov chain Monte Carlo (MCMC) method using 10\,000 iterations for each flare. In addition, the vertical green dashed line indicates the date of the main flare center (also listed in Table~\ref{table:flare_characteristics} with its uncertainty).

The periods between the flares are computed as the time elapsed between these main flare centers. As a result, $P_{1\rightarrow2} = (3.20 \pm 0.13)$ yr is obtained for the time between main flare one and two, $P_{2\rightarrow3} = (3.07 \pm 0.09)$ yr between two and three and $P_{3\rightarrow4} = (3.06 \pm 0.17)$ yr between three and four (see Table~\ref{table:flare_characteristics}).

\section{Jet kinematics at 8.6 GHz, utilizing archival VLBI data}

To derive the structural and kinematic properties of the mas-scale jet of J1048+7143, we used archival calibrated $8.6$-GHz very long baseline interferometry (VLBI) visibility data taken with the Very Long Baseline Array (VLBA). The observations span more than $26$~yr at 69 epochs (between 1994.61 and 2020.73), and the calibrated visibilities are available in the Astrogeo database\footnote{\url{http://astrogeo.org/cgi-bin/imdb_get_source.csh?source=J1048\%2B7143}}. We introduced these observations already in \citet{Kun2022a}.

We model the brightness distribution of the source visibility data by fitting elliptical Gaussian components \citep{Pearson1995} to the core and circular Gaussians to the jet components using the \textsc {Difmap} \citep{Shepherd1994} software. The fitted parameters are the component flux density, the position, the component width, and for only the core the position angle of the major axis (measured from north through east) and the minor-to-major axial ratio of this elliptical component. We followed \citet{Kun2014} for the error estimation of the fitted parameters. 

For VLBI jets, seen in small viewing or inclination angle with respect to the line of sight, the apparent brightness temperature $T_\mathrm{b}$ usually exceeds the limiting intrinsic brightness temperature $T_\mathrm{int}$ due to strong Doppler boosting. The Doppler factor $\delta$ connects them as $T_\mathrm{b}=\delta T_\mathrm{int}$. The brightness temperature of the VLBI components is estimated as \citep[e.g.][]{Condon1982}:
\begin{equation}
    T_\mathrm{b,VLBI}=1.22\cdot 10^{12}\times(1+z) \frac{S_{\nu}}{d_1 d_2 f^2} \mbox{~}\mathrm{(K)},
\end{equation}
where $S_{\nu}$ is the flux density (in Jy), $d_1$ and $d_2$ are the major and minor axes of the core (in mas), respectively, $f$ is the observing frequency (in GHz), and $z=1.15$ \citep{Polatidis} is the redshift of the source. 
Assuming that the intrinsic brightness temperature is slightly lower than the equipartition brightness temperature, $T_\mathrm{int} \approx 3\times10^{10}$~K \citep{Homan2006}, we obtained the median value of $\delta\approx5.1$ for the core at 8.6 GHz.

We were able to identify two moving components, C1 and C2. We calculated the apparent proper motion of the components first by fitting linear proper motions as
\begin{equation}
r_\mathrm{L}=\mu (t-t_\mathrm{ej}),
\label{eq:betalin}
\end{equation}
and then by fitting accelerating proper motions as 
\begin{equation}
r_\mathrm{A}=r_\mathrm{L} + \frac{\dot{\mu}}{2} \left( t-t_\mathrm{mid}\right)^2,
\label{eq:betaapp}
\end{equation}
where $\mu$ is the linear proper motion measured in mas\,yr$^{-1}$, $t_\mathrm{ej}$ the ejection time, $\dot{\mu}$ the angular acceleration, and $t_\mathrm{mid}$ the half of the sum of the maximum and the minimum epochs when the respective components are detected. 

\begin{table*}
\centering
\caption{Component speed fits. (1) jet component identifier. From linear fit (L): (2) linear proper motion, (3) ejection time, (4) reduced $\chi^2$. From accelerated fit (A): (5) non-linear proper motion, (6) acceleration rate, (7) ejection time, (8) reduced $\chi^2$. }
\begin{tabular}{ccccccccc}
\hline
\hline
ID$^{(1)}$ & $\mu_\mathrm{L}^{(2)}$ & $t_\mathrm{ej,L}^{(3)}$ & ${\chi^2}^{(4)}$& $\mu_\mathrm{A}^{(5)}$ & ${\dot{\mu}_\mathrm{A}}^{(6)}$ & $t_\mathrm{ej,A}^{(7)}$ & ${\chi^2}^{(8)}$ \\ 
 & (mas yr$^{-1}$) & (yr) & & (mas yr$^{-1}$) & (mas yr$^{-2}$) & (yr) & & \\ 
\hline
C1 & $0.118\pm0.006$ & $1996.160\pm0.250$ & 3.503 & $0.113\pm0.002$ & $-0.033\pm0.004$ & $1995.676\pm0.139$ &0.578\\
C2 & $0.074\pm0.016$ & $1993.811\pm3.182$ & 5.034 & $0.073\pm0.014$ & $0.020\pm0.011$ & $1994.565\pm2.606$ &3.006\\
 \hline
\hline
\label{table:velocityfits}
\end{tabular}
\end{table*}

\begin{figure*}
    \centering
    \includegraphics[scale=0.48]{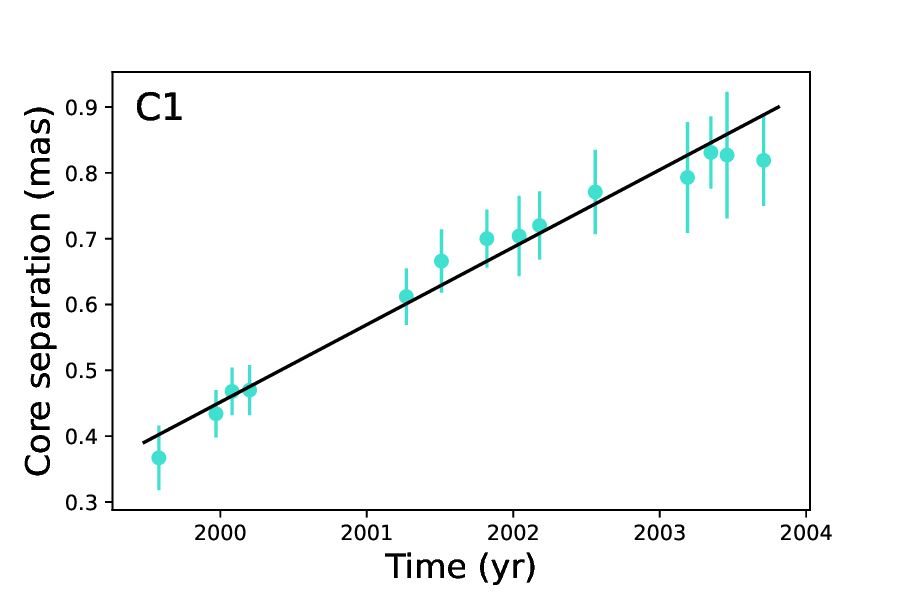}
    \includegraphics[scale=0.48]{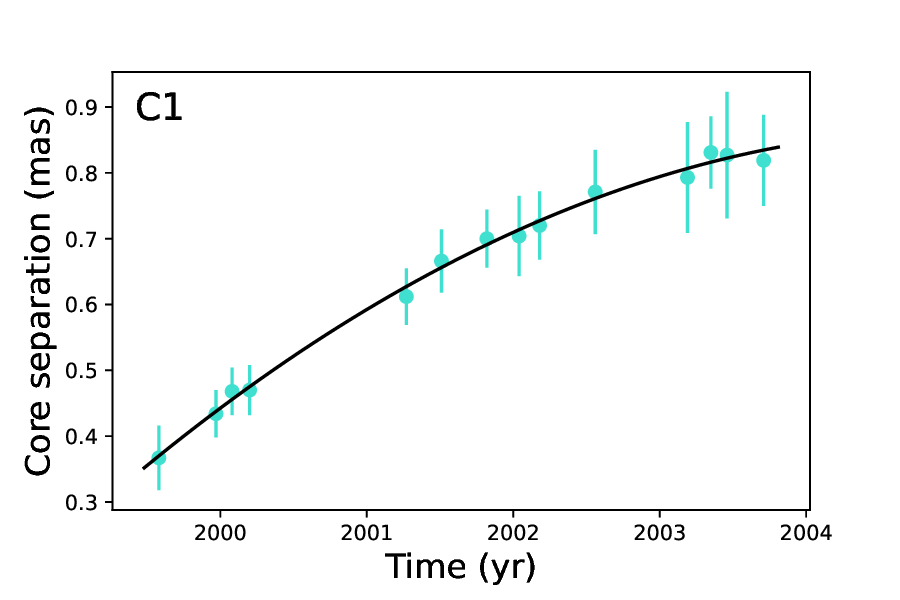}\\
    \includegraphics[scale=0.48]{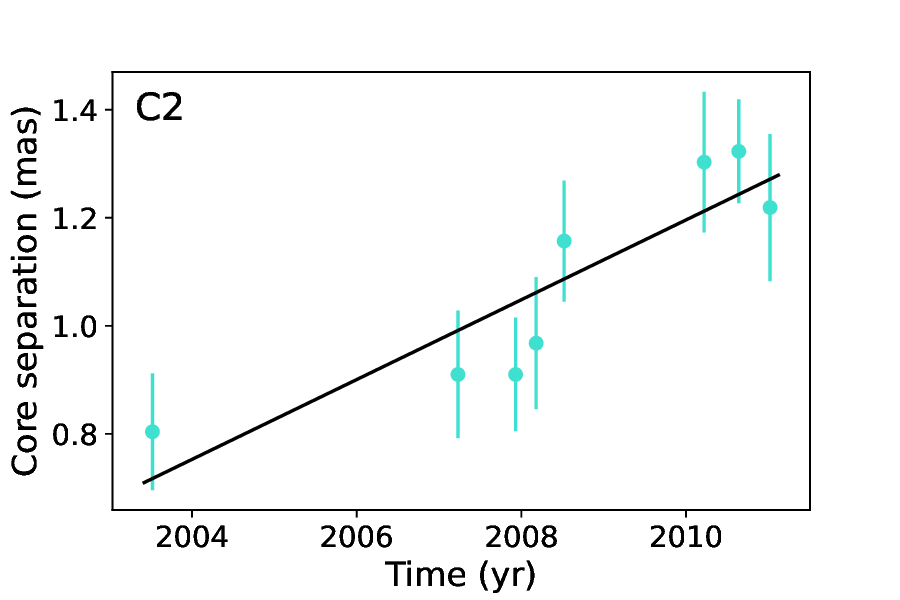}
    \includegraphics[scale=0.48]{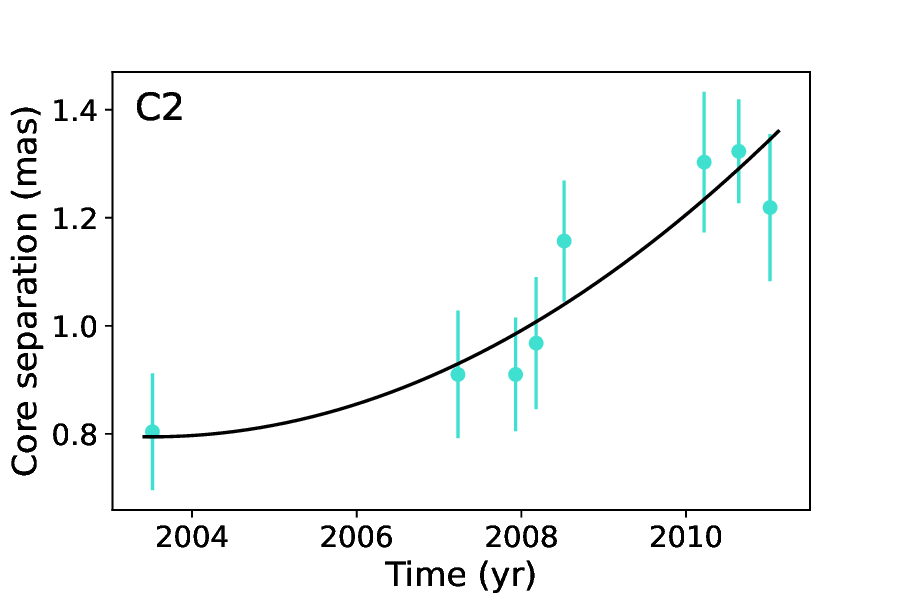}
    \caption{Linear (left) and accelerated (right) proper motion fits for jet components C1 and C2 of J1048+7143 at $8.6$~GHz. Only observations before the flaring behavior of J1048+7143 started are used to make these plots.}
    \label{fig:propmotion}
\end{figure*}

We fitted the core separation of C1 and C2 as function of time before the flaring behavior of J1048+7143 started according to its single-dish radio flux density curve \citep[see in][]{Kun2022a}. We show the proper motion fits in Fig.~\ref{fig:propmotion}. The resulted best-fit proper motions, as well as the derived apparent speeds are shown in Table~\ref{table:velocityfits}. As it can be seen from the reduced $\chi^2$ values in Table~\ref{table:velocityfits} and from Fig.~\ref{fig:propmotion}, the accelerated proper motion fits give better results. 

The apparent speed of the jet components is related to the bulk jet speed $\beta$ and their inclination angle $\iota$ as \citep[e.g.][]{Urry1995}
\begin{equation}
\beta_\mathrm{app}=\frac{\beta \sin \iota}{1-\beta \cos \iota},
\end{equation}
where the jet speed $\beta$ is related to the Lorentz factor $\Gamma$ as
\begin{equation}
\beta=\sqrt{1-\frac{1}{\Gamma^2}}.
\end{equation}
The maximal $\beta_\mathrm{app}$ occurs for the critical inclination $\cos \iota_c=\beta$, $\sin \iota_c=\Gamma^{-1}$, having the value $\beta_\mathrm{app,max}=(\Gamma^2-1)^{1/2}$.

In practice, it is unlikely that a jet component will exactly move in this direction, therefore the maximal speed $ \beta_\mathrm{app,max} ^\mathrm{obs}\leq\beta_\mathrm{app,max}$ seen in the VLBI jet gives a lower limit on the Lorentz factor of the jet
\begin{equation}
\Gamma_\mathrm{low}=\sqrt{1+(\beta_\mathrm{app,max}^\mathrm{obs})^2}.
\label{eq:gammamin}
\end{equation}

Taking the accelerated proper motion of C1, $\mu_\mathrm{max}=(0.113\pm0.002)$~mas\,yr$^{-1}$, that is the fastest jet component seen in J1048+7143 before the flaring behavior starts, we get the maximum observed apparent speed as $\beta_\mathrm{app,max} ^\mathrm{obs}=6.56\pm0.12$. Then the limiting Lorentz factor emerges as $\Gamma_\mathrm{low}=6.87$ and the minimum intrinsic jet speed is $\beta_\mathrm{low}=0.989$ (in the units of the speed of the light). Then the related critical inclination angle is $\iota_\mathrm{c}=\arcsin (\Gamma_\mathrm{low}^{-1})\approx 10.9\degr$ and the half-opening angle of the jet is estimated as $\zeta\approx 1/\Gamma\approx 8.3^\circ$ \citep{Boettcher2012}.

\section{Spin--orbit Precession and Prediction of the Next $\gamma$-ray Flare}
\label{sec:predictFlare}

\begin{figure*}
    \centering
    \includegraphics[angle=0,scale=0.425]{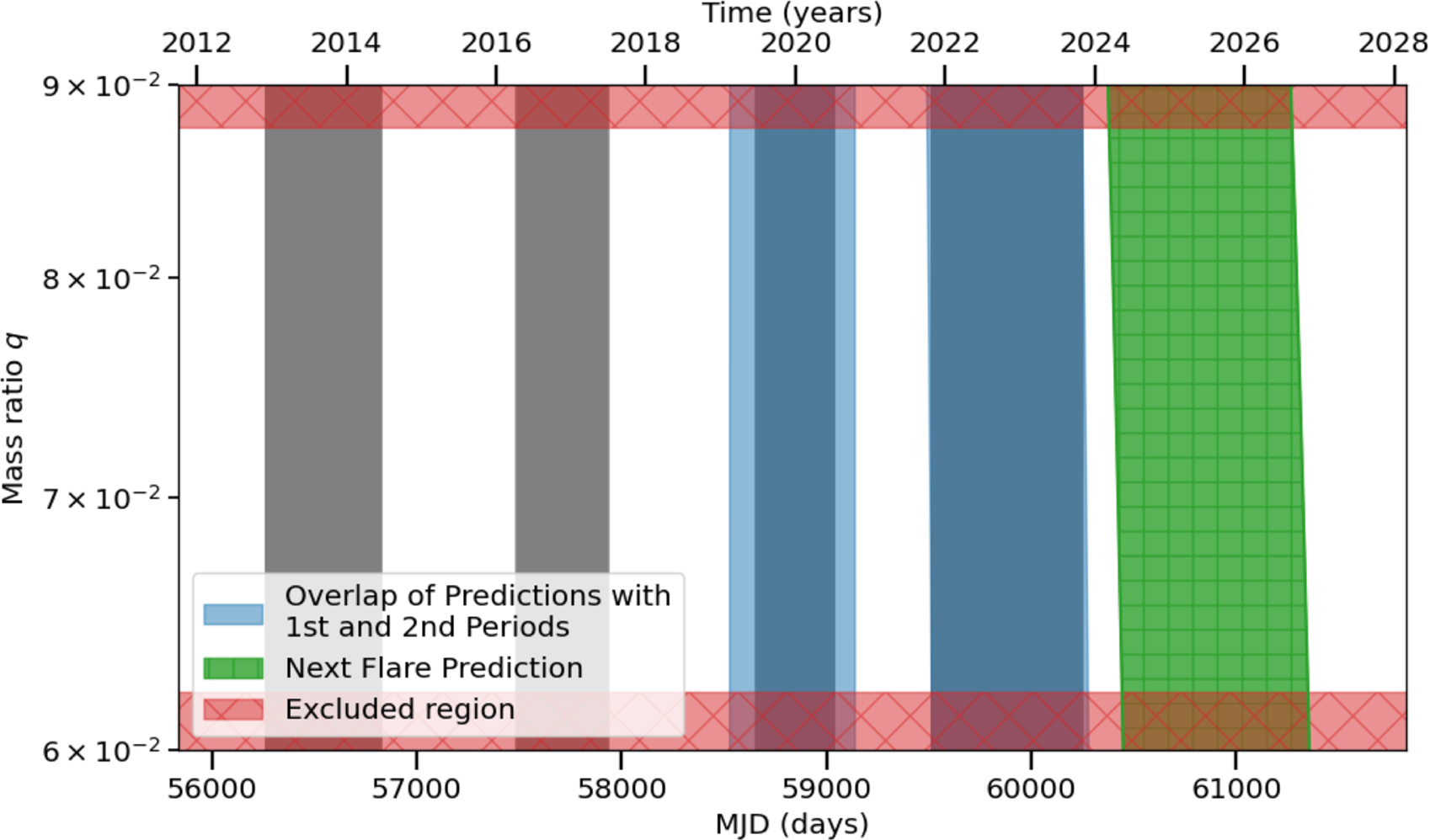}
    \caption{Prediction plot for $\gamma$-ray flares from J1048+7143. Gray areas show the duration's of the four flares determined with the centroid method in Table~\ref{table:flare_characteristics}. 
    The blue hashed bands represent an overlap of the predictions using the period $P_{n\rightarrow n+1}$, where $n=1$ or $2$. Invalid mass ratios are displayed via the red crossed area. This plot is based on the duration of the second flare and the jet half-opening angle $\zeta=8.3^\circ$. The latter was derived from the pc-scale jet kinematics of J1048+7143 seen at the $f=8.6$~GHz observing frequency with VLBA.}
    \label{fig:flarePredict}
\end{figure*}

In \citet{Kun2022a} we have shown that the $\gamma$-ray light curve  of J1048+7143 ($E>168$ MeV) and its single-dish radio flux density curve obtained at $4.8$~GHz (see bottom panel of Fig. \ref{fig:centroid_Method}) are consistent with the jet precession driven by the spin--orbit precession, occurring in a SMBBH with a total mass of $m=10^{9.16}M_\odot$. We were able to constrain the mass ratio only from the above, such that $q<0.2$, where $q=m_2/m_1$ ($m=m_1+m_2$, $m_1>m_2$). We note, the $\sim 0.1$ Jy difference in flux density seen between the single dish radio flux density curves in Fig. \ref{fig:centroid_Method} is most probably caused by the difference in the resolving power between the RATAN600 and Nanshan (Urumqi) radio telescopes at 4.8 GHz.

In the inspiral phase of the merger of a SMBBH, first the dynamical friction, then the gravitational radiation shrinks the orbit of the black holes. In the latter case, the binary goes through three phases, the inspiral, the plunge, and the ring-down. In the inspiral phase, the equation of motion can be approximated by a series expansion in terms of a small parameter, the post-Newtonian (PN) parameter $\varepsilon=Gm (c^2 r)^{-1}$, where $G$ is the gravitational constant, $c$ is the speed of the light, $m$ is the total mass, and $r$ is the separation of the binary.

The compact binary dynamics is conservative up to the second PN order, such that the total energy $E$ and the total momentum $\mathbf{J}=\mathbf{S}_1+\mathbf{S}_2+\mathbf{L_N}$ are conserved during the motion. If the $\mathbf{S}_1$ and $\mathbf{S}_2$ the spins of the two SMBHs ($m_1>m_2$) are not parallel with the Newtonian orbital momentum $\mathbf{L_N}$, the spins begin to precess \citep{Barker1975,Barker1979}:
\begin{equation}
\mathbf{\dot{S}}_i\!=\!\mathbf{\Omega}_i \times \mathbf{S}_i,
\end{equation}
where $\mathbf{\Omega}_i$ is the angular momentum of the $i$th spin and it contains spin--orbit (1.5PN), spin--spin (2PN) and quadrupole--monopole contributions (2PN) up to 2PN order. The gravitational radiation circularizes the orbits \citep{Peters1964}, therefore from here we take only circular orbits into account.

Expanding the work of \citet{deBruijn2020}, in \citet{Kun2022a} we determined the direction angle of the dominant spin ($\phi$) as a function till the final coalescence of the compact binary as:
\begin{eqnarray}
    &&\phi(\Delta T_{\rm GW} \,,\, q) = - \frac{2 \, (4 + 3q)}{(1 + q)^2}\times \nonumber \\
           &&\, \times \left( \frac{5 \, c}{32 \,G^{1/3} m^{1/3}} \cdot \frac{(1 + q)^2}{q} \right)^{3/4} \left(\Delta T_{\rm GW}\right)^{1/4} 
           + \psi,
           \label{eq:JetModel}
\end{eqnarray}
where $\psi$ is an integration constant.

Assuming the dominant spin makes a complete $360^\circ$ turn between the subsequent major flares in the $\gamma$-ray light curve, we can establish the connection between two flares as:
\begin{equation}
    \phi(\Delta T_{\rm GW} \,,\, q) = \phi(\Delta T_{\rm GW} - P_{\rm jet} \,,\, q) \pm \zeta \,
    \label{eq:determine_T_GW},
\end{equation}
where $P_\mathrm{jet}$ is the jet precession period, which is the time that elapsed between two subsequent major flares in the $\gamma$-ray light curve. 

Updating the light curve with the 4th major $\gamma$-ray flare, the mass ratio of the SMBBH is now constrained as $0.062<q< 0.088$ (see Fig.~\ref{fig:flarePredict}). In our previous work, we were able to constrain the mass ratio as $q<1/4$. This means that including the new flare, now we can constrain the mass ratio into a significantly narrower range, $0.062<q<0.088$, and consequently we can further constrain the parameters of the hypothetical SMBBH at the heart of J1048+7143. Since the emission of gravitational waves leads to shrinking orbits, the Newtonian orbital angular momentum $\mathbf{L_N}$ and the direction of the dominant spin $\mathbf{S_1}$ also change: the former is moving away, while the latter is moving closer to the total angular momentum $\mathbf{J}$. Therefore the jet, attached to $\mathbf{S_1}$, also shows a slow, secular change in its direction. We see the Doppler boosting governed by the spin--orbit precession while the jet lies close to our line of sight. Once the jet is far away from our line of sight because of the secular change in the direction of $\mathbf{S_1}$, we do not expect a flaring behavior anymore due to the weak Doppler boosting. Given the jet still boosted in the close future, a 5th $\gamma$-ray flare is expected to arrive between MJD 60379 (2024 March 10) and MJD 61350 (2026 Nov 6).

\section{Discussion and Conclusions}
\label{sec:sumconcl}
\begin{figure}
    \centering
    \includegraphics[scale=0.275]{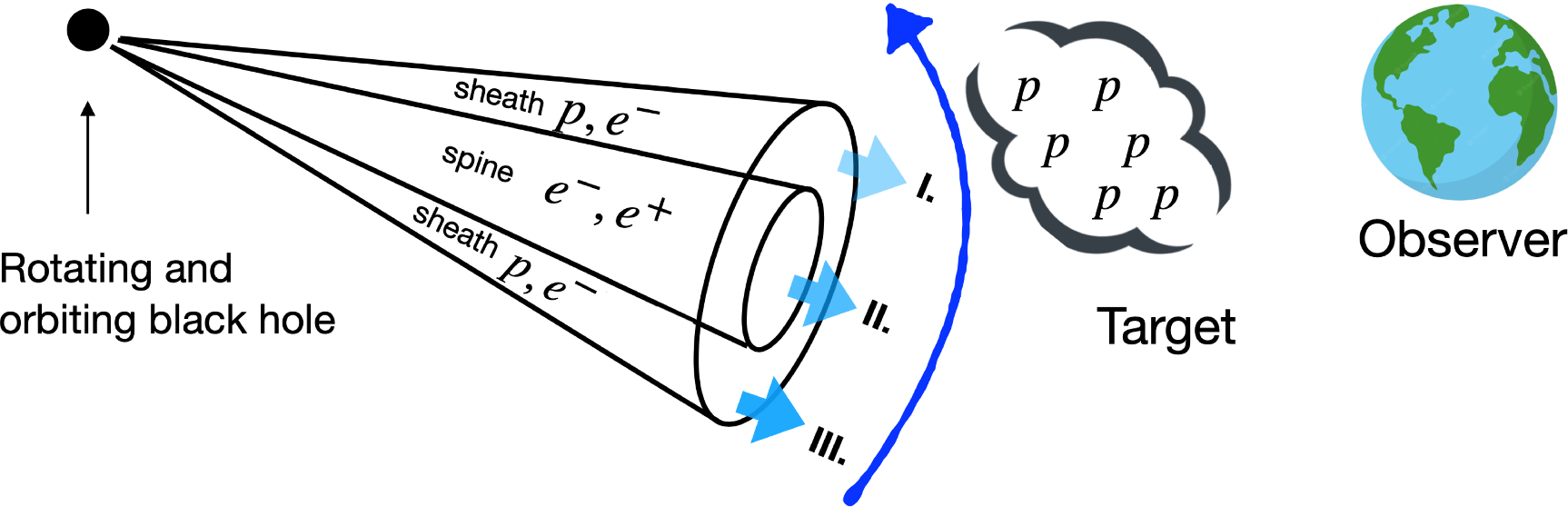}
    \caption{Geometry of the jet-precession-induced light curve variation. The jet, driven by the spin--orbit precession in a SMBBH, goes through a target that supplies target protons for the hadronic processes between them and high-energy cosmic rays accelerated in spine--sheath structured jet.}
    \label{fig:sketch}
\end{figure}

Periodicites in AGN can usually be explained by a number of model families, relying on single \citep[e.g.][]{2003ApJ...591L.119L} or binary black holes \citep[e.g.][]{1993ApJ...409..130R,2012ApJ...759..118B,2018MNRAS.478..359K,2019ApJ...873...11J,2020MNRAS.496.3336K,2023MNRAS.526.4698K,britzen23}, on the accretion disk \citep{Bardeen1975,2006ApJ...638..120C}, or instabilities within the jet \citep[e.g.][]{1994ApJ...424..126H,2006A&A...456..493P}. Application of these models depend on the complexity of the observed behavior: while models relying on e.g. single black holes or the precession of accretion disk can usually explain a stable period, for cases when the time duration between flares changes, inclusions are needed. We note that in quasars, quasi-periodic oscillations (QPOs) can be attributed to stochastic changes in the accretion disk \citep[e.g.][]{2022ApJS..263...16K}, or disk disturbances due to companion black holes \citep[e.g.][]{2004MNRAS.349.1218C} etc. However, J1048+7143 belongs to the blazar subclass of radio-loud AGN as its jet is seen under a small viewing angle. For such sources, relativistic beaming dominates the electromagnetic emission as the boosted jet can overshine any intrinsic signal.

Here we offer a qualitative scenario to explain the multiwavelength behavior of J1048+7143 (see the $\gamma$-ray light curve and single-dish radio flux density curve in Fig.~\ref{fig:centroid_Method}). We note that any modeling targeting this source should explain the decreasing nature of the time duration between subsequent flares and their double structure in $\gamma$-rays. The scenario we found to be the most compatible with these is the spin--orbit precession of the supermassive black hole that launches a spine--sheath jet and drives the multiwavelength appearance of the source. Due to the precession, an ($e^-$, $e^+$) spine -- ($p$, $e^-$) sheath jet periodically goes through our line of sight. The precessing jet approaches the target (e.g. clouds in the broad-line region) close to our line of sight. Their interaction imprints 3 phases in the $\gamma$-ray and radio regimes. First the ``upper'' sheath with accelerated cosmic rays meets the target (that gives the target protons), which leads to the first (pionic) $\gamma$-ray subflare. Meanwhile, the radio flux density goes up as relativistic boosting gets stronger due to the decreasing viewing angle (Phase I in Fig.~\ref{fig:sketch}). Then the spine meets the target leading to the minimum in the pionic $\gamma$-rays between the subflares since the proton density in the spine is negligible and therefore there are no hadronic interactions between the jet and the target (Phase II in Fig.~\ref{fig:sketch}). Meanwhile, the radio flux density curve reaches its maximum as we see the jet at the smallest viewing angle. This boosted radio emission is expected to come dominantly from the spine of the jet. Then the ``lower'' sheath meets the target that leads to the second $\gamma$-ray subflare, while the radio flux density goes down as relativistic boosting gets weaker due to the increasing viewing angle of the jet (Phase III in Fig.~\ref{fig:sketch}). Then the jet leaves the target and moves more away from our line of sight. Since the VLBI jet is visible between the major flares, hinting that considerable boosting is still present, we can conclude that the jet moves not far from our line of sight. This is when the quiescence phase happens between the major flares. After one spin--orbit period, the jet approaches again, triggering the next sequence of I/II/III phases (Fig.~\ref{fig:sketch}). 
We note that the optical light curve of J1048+7143, constructed based on visual optical observations sent to the American Association of Variable Star Observers (AAVSO), also show the double flaring structure, similarly to the $\gamma$-ray light curve. A follow-up paper elaborating on the optical light curves of J1048+7143 is in preparation.

Due to gravitational radiation, the dominant spin $\mathbf{S_1}$ slowly approaches the total angular momentum $\mathbf{J}$, therefore the jet meets the target at different angles as the binary merger is progressing. After several spin--orbit periods, the jet just does not meet with the target again, as the precession cone of $\mathbf{S_1}$ is narrowing. The $\sim90^\circ$ misalignment between the pc-scale and kpc-scale jets of J1048+7143 is indeed compatible with a secular change in the jet direction \citep{Kun2022a}. Qualitatively it explains why we do not see a flaring radio behavior before 2008 -- the jet is circling somewhere away from our line of sight. The peak flux of the $\gamma$-ray subflares is first increasing then decreasing. It can also be explained in the framework of this scenario, with the combination of changing characteristic relativistic boosting of the jet and column density changes in the target.

Since in phases I and III the emission of high-energy neutrinos is expected from $p-p$ interactions, J1048+7143 might be an interesting source to search for high-energy neutrinos in archival neutrino datasets. We encourage further multiwavelength and multimessenger studies of this source.

\begin{acknowledgments}
E.K. thanks the Alexander von Humboldt Foundation for its Fellowship. J.T. and I.J. acknowledges support from the German Science Foundation DFG, via the Collaborative Research Center \textit{SFB1491: Cosmic Interacting Matters -- from Source to Signal}. Support from the Hungarian National Research, Development and Innovation Office (NKFIH) is acknowledged (grant number OTKA K134213). This project has received funding from the HUN-REN Hungarian Research Network. L.C. was supported by the CAS ``Light of West China'' Program (No. 2021-XBQNXZ-005) and the Xinjiang Tianshan Talent Program. 
This paper makes use of publicly available \textit{Fermi}-LAT data provided online by the \url{https://fermi.gsfc.nasa.gov/ssc/data/access/} Fermi Science Support Center. On behalf of Project `fermi-agn', we thank for the usage of the HUN-REN Cloud that significantly helped us achieving the results published in this paper. The National Radio Astronomy Observatory is a facility of the National Science Foundation operated under the cooperative agreement by Associated Universities, Inc. The XAO-NSRT is operated by the Urumqi Nanshan Astronomy and Deep Space Exploration Observation and Research Station of Xinjiang (XJYWZ2303). We acknowledge the use of data from the Astrogeo Center database maintained by Leonid Petrov.
\end{acknowledgments}

\end{document}